\documentclass[fleqn,usenatbib]{mnras}

\usepackage{graphicx}	
\usepackage{amsmath}	
\usepackage{amssymb}	

\usepackage[T1]{fontenc}
\usepackage{ae,aecompl}

\usepackage{newtxtext,newtxmath}

\def\aap{A\&A}
\def\apj{ApJ}

\def\apjl{ApJ}
\def\mnras{MNRAS}

\def\apjs{ApJS}

\def\ltsima{$\; \buildrel < \over \sim \;$}
\def\simlt{\lower.5ex\hbox{\ltsima}}
\def\gtsima{$\; \buildrel > \over \sim \;$}
\def\simgt{\lower.5ex\hbox{\gtsima}}

  \newcommand{\bc}{\begin{center}}
  \newcommand{\ec}{\end{center}}
  
  \newcommand{\hMsun}{~h^{-1}\>{\rm M_\odot}}
  \newcommand{\Mpc}{~h^{-1}~{\rm Mpc}}
  \newcommand{\Kpc}{~h^{-1}~{\rm kpc}}

\let\ACMmaketitle=\maketitle
\renewcommand{\maketitle}{\begingroup\let\footnote=\thanks \ACMmaketitle\endgroup}
\title[Trends in Modern Cosmology]{The Impact of Baryons on the Large-Scale Structure of the Universe\footnote{Original PDF file for this chapter: https://cdn.intechopen.com/pdfs-wm/54705.pdf}}
\author[W. Cui and Y. Zhang]
{\parbox{\textwidth}{Weiguang Cui$^{1}$\thanks{E-mail: \texttt{cuiweiguang@gmail.com}}
 and Youcai Zhang$^{2}$.
 }\vspace{0.4cm}
 \\
 \parbox{\textwidth}{
$^1$ Dover Court 4/88, Western Australia, Australia\\
$^2$ ShangHai Astronomical Observatory, Shanghai, China
}}

\date{http://dx.doi.org/10.5772/68116}

\pubyear{2017}

\begin{document}
\label{firstpage}
\pagerange{\pageref{firstpage}--\pageref{lastpage}}

\maketitle
\begin{abstract}
Numerical simulations play an important role in current astronomy researches. Previous
dark-matter-only simulations have represented the large-scale structure of the Universe.
However, nowadays, hydro-dynamical simulations with baryonic models, which can
directly present realistic galaxies, may twist these results from dark-matter-only simulations.
In this chapter, we mainly focus on these three statistical methods: power spectrum,
two-point correlation function and halo mass function, which are normally used to
characterize the large-scale structure of the Universe. We review how these baryon processes
influence the cosmology structures from very large scale to quasi-linear and non-linear scales by comparing dark-matter-only simulations with their hydro-dynamical
counterparts. At last, we make a brief discussion on the impacts coming from different
baryon models and simulation codes.
\end{abstract}

\begin{keywords}
  cosmology: large-scale structure, simulation, statistical methods, hydro-dynamical
  simulation, baryonic models
\end{keywords}


\section{Introduction}
\label{i}
The core of current research foci in cosmology is to interpret the distribution and properties of observed galaxies in the sky and to understand their formation and evolution. The current standard cosmology model -- lambda-cold-dark-matter ($\Lambda$CDM) paradigm -- provides a general explanation for the galaxy formation and evolution: matter is dominated by the dark matter, which only subjects to gravitational interactions; inside dark matter halo that acts as a gravitational potential well, baryonic matters go through a series of physical processes, such as gas cooling, star forming and death with Supernova feedback \citep[for example][]{Mo2010}.

From the cosmic micro background (CMB) observation, such as WMAP \citep{Spergel2003} and Plank \citep{PlanckCollaboration2013}, matters occupy
roughly one-fourth of total energy of the Universe. The rest comes from dark energy. Dark matter is
about $20\%$ of the total energy, while baryons only occupy $5\%$ \citep[see more accurate fractions from][]{PlanckCollaboration2015}. At the CMB time (z $\sim$ 1100), matters are distributed nearly 'homogeneous' in the Universe with little fluctuations at small scales. Started from that time, dark matter and baryons are assembled by the gravitational force. They follow a pattern of hierarchical structure formation, where the smaller structures form first, then merge to build massive ones \citep[see for example][for details]{Planelles2014a}. At very large scale, this structure formation process can be roughly described by the Zel'dovich approximation \citep{Zeldovich1970}. However, the formation of structures with gravity is a nonlinear process, which cannot be fully described analytically, especially at small scales. Therefore, building these structures and tracing their evolutions require numerically solving the gravitational equation.

Combining with modern computers, this problem can be solved with numerical methods  -- N-body simulations, which boosts a new area of research in astronomy. Initially, different numerical methods are developed to simulate only dark matter component, such as particle-mesh (PM), particle-particle/particle-mesh (P3M) and tree-PM algorithms. Dark matter is described numerically by data points/particles that trace a mass element corresponding to a volume element of the early 'homogeneous' universe. Those methods successfully describe the formation of structures by implementing gravitational interactions. Thus, over decades, such simulations have been widely used with little variation in term of physics. Combined with ever decreasing limitations of computer resources and vast improvement in terms of implementations, larger volumes can be explored with increasing resolution to reserve the small-scale information. The properties of cosmology structures (such as cosmic web, voids), halos and even subhaloes are well understood.

{\em However, these simulations cannot directly give any information of galaxies, which are resident inside dark matter halos.}

To connect these theoretical investigation results with observed galaxies, numerous methods are developed. They can be roughly separated into these three approaches:

\begin{itemize}
  \item Much simpler approaches are halo occupation distribution (HOD) models, where observed galaxies are assigned to halos by matching both the halo mass and stellar mass functions \citep[e.g.][]{Jing1998, Jing2002, Yang2003, Bosch2007, Rodriguez-Puebla2015, Zu2015}. Such methods are tuned to directly link the luminosity functions with halo mass functions. Thus, they are successful in defining the stellar mass halo mass (SMHM) relation. However, this method cannot provide useful individual galaxy information. Furthermore, the scatter in this relation still remains uncertain and difficult to interpret. It can be constrained by comparing specific galaxies, their environments, the inter galactic medium (IGM) and their full formation history.
  \item Other less computationally intensive methods involve applying sub-grid models on the scale of dark matter halos, starting from the accretion of gas by the potential well, following recipes of gas cooling, star forming, supernova (SN) and active galactic nuclei (AGN) feedbacks, at last galaxies are formed and evolved under the halo merger tree. These semi-analytical models (SAMs) have been successfully applied to halo catalogues extracted from N-body simulations \citep[e.g.][]{White1991, Mo1996, DeLucia2004, Croton2005, Kang2005, Baugh2006, Guo2010}. Interested readers are encouraging to find the differences between these models (including HOD models) in the nIFTy cosmology comparison project \citep{Knebe2015, Pujol2017} and their following works. As the formation of halos can be traced in the form of halo merger trees, both the formation and interaction of galaxies can be explored within the time frame and the mass resolution explored by the simulation. Although these methods can provide more physical views of galaxy formation, they are still lacking the consistency of co-evolving between baryon and dark matter.
  \item Hydro-dynamical simulations are the only way to overcome the problem faced by SAMs. They can directly solve the physical processes of the baryonic component on top of the dark matter one, which can provide consistent co-evolution with the same gravitational force. These hydro-simulations require complex implementations of baryonic models with gas described either as (a) numerical data points with associated density (smooth particles hydrodynamics (SPH): \cite{Springel2001a,Springel2005,Wadsley2003,Beck2016}, etc.), (b) grid cells fixed in the volume (cells are refined and unrefined as required to explore highest gas density while neglecting low-density regions with nested mesh or adaptive mesh refinement (AMR): \cite{Kravtsov1997,Teyssier2002,TheEnzoCollaboration2014}, etc.) or (c) moving mesh (the gas element is associated with a numerical point within a volume defined from the distribution of nearby mesh point through Voronoi tessellation \citep{Springel2009}). The key aspect is the description of the physical processes within these gas elements. These recipes from SAM can be implanted in hydro-dynamical simulations with moderate modifications. However, hydro-dynamical simulation is suffered from its time-consuming computation, with which the numerous free parameters from these sub-grid baryonic models cannot be easily tuned to represent these observational relations as they are in SAM.
\end{itemize}

Although hydro-dynamical simulations are the heaviest and most time-consuming tool for connecting the dark part with the luminous part in the Universe, they are irreplaceable in investigating/understanding galaxy formations in a full picture. Those HOD and SAM models, which are used to create mock galaxy catalogues, have been quite successfully in reproducing the observational statistical features, such as the two-point correlation functions, luminosity functions, colour distributions and star formation rates. Nevertheless, they are based on the assumption that baryon processes are independent of dark matter halo formation, which is apparently not true \citep{Lewis2000,Gnedin2004,Lin2006}. As both observation and simulation are becoming more and more accurate, the back reaction of baryons to dark matter cannot be ignored. Thanks to the Morse's law, more and more efforts are being put in these areas in recent years, for example, \cite{Cui2012a,Cui2014a}, the OWLS project \citep{Schaye2009}, the EAGLE project \citep{Schaye2015}, the Illustris project \citep{Vogelsberger2014} and the Horizon-AGN simulation \citep{Dubois2016} for these cosmological simulations; \cite{Planelles2013,Planelles2014}, the NIHAO project \citep{Wang2015} and the FIRE project \citep{Hopkins2014} for these zoom-in simulations. Interested readers refer to the Aquila project \citep{Scannapieco2012}, the AGORA project \citep{Kim2013} and the nIFTy cluster comparison project \citep{Sembolini2016,Sembolini2016a,Elahi2016,Cui2016,Arthur2017} for the comparison of different hydro-dynamical simulation codes. A number of studies based on cosmological hydro-dynamic simulations have been recently carried out to analyse in detail the effect of baryonic processes on different properties of the total mass distribution, such as the power spectrum of matter density fluctuations \citep[e.g.][]{Rudd2008,VanDaalen2011,Casarini2012,VanDaalen2015}, the halo correlation functions \citep{Zhu2012,VanDaalen2014}, the halo density profiles \citep[e.g.][]{Lin2006,Duffy2010,Killedar2012,Ragone-Figueroa2012,Schaller2015}, concentration \citep[e.g.][]{Bhattacharya2013,Rasia2013}, halo shape \citep[e.g.][]{Knebe2010,Cui2016}, dynamical state \citep[e.g.]{Cui2017a} and the (sub-) halo mass function \citep[e.g.][]{Stanek2008,Cui2012a,Sawala2013,Martizzi2014,Cusworth2014,Cui2014a,Velliscig2014,Despali2017}.

In this chapter, we will focus on the impacts of baryons through these comparisons between hydro-dynamical simulations with dark-matter-only simulations and summarize the results in these three aspects: power spectrum, two-point correlation function (2-PCF) and halo mass function (HMF).

\section{Chapter}

In the last decade, dark-matter-only simulations have been vastly used to theoretically investigate the large-scale structure of the Universe. Through different statistical methods, such as power spectrum, two-point correlation function, halo mass function and so on, the formation and evolution of the large-scale structures have been clearly characterized by those cosmological dark-matter-only simulations. However, the observed Universe can only show the distribution of baryonic matters at such scales. To connect these theoretical understanding with observations of the large-scale structure of the Universe, we need hydro-dynamical simulations, which can provide a consistent evolution driving by the gravitational force for both dark matter and baryons. With these hydro-simulations, we can directly compare simulations with observations through mock techniques \citep[e.g.][]{Cui2011,Cui2014b,Cui2016a}; explore the galaxy formation process in details; correct and improve our understanding of these baryon models, and so on. In this chapter, we only concentrate on one simple question: How do the baryon processes react on dark matter? This is a question, which these simplified analytical models such as HOD and SAM with ad hoc parameters lack the ability to deal with. As baryons occupy only a small fraction of total matter, we are expecting a very weak effect on the dark matter structures. Nevertheless, baryons dominate at small scales such as in galaxies, where the effect cannot be ignored anymore. Thus, we will address this question with different statistical quantities at different scales, which are listed in Sections 2.1, 2.2 and 2.3.

\subsection{POWER SPECTRUM}

The power spectrum P(k) (here k is the co-moving wavenumber corresponding to a co-moving spatial scale $\lambda=2\pi/k$) is one of the most powerful and basic statistical measurements that describes the distribution of mass in the Universe, and one of the most thoroughly investigated quantities in modelling the structure formation process. Due to the large amount of data from both observation and simulation, the power spectra are measured mostly using the fast Fourier transform (FFT) technique. Lots of methods are used to improve the accuracy of the measurement for power spectrum especially at nonlinear scale \citep[for example][]{Cui2008,Colombi2008}. However, such algorithm improvements cannot deal with the power spectrum changes caused by the physical models.

Using the OWSL simulations, \cite{VanDaalen2011} studied the influence of baryonic models on matter power spectrum through a comparison between a dark-matter-only (DMONLY) one and hydro-dynamical simulations (REF and AGN). Starting from the same initial condition, these simulations from various models are listed in Table 1.

\begin{table*}
 \label{tab:continued}
 \begin{tabular}{ll}
  \hline
  Simulation & Description \\
  REF & Reference simulation, includes radiative cooling and heating, star forming with the Chabrier (2003) \\
      & stellar initial mass function and SN feedback with wind mass loading $\eta = 2$ and velocity $v_w = 600 km~ s^{-1}$ \\
  AGN & Includes AGN (in addition to SN feedback)  \\
  DMONLY & No baryons, CDM only  \\
  \hline
\end{tabular}
\caption{Different variations on the reference simulation that are compared in the chapter. Unless noted otherwise, all simulations use a set of cosmological parameters derived from the WMAP3 results and use identical initial conditions.}
\end{table*}

\begin{figure*}
 \includegraphics[width=\textwidth]{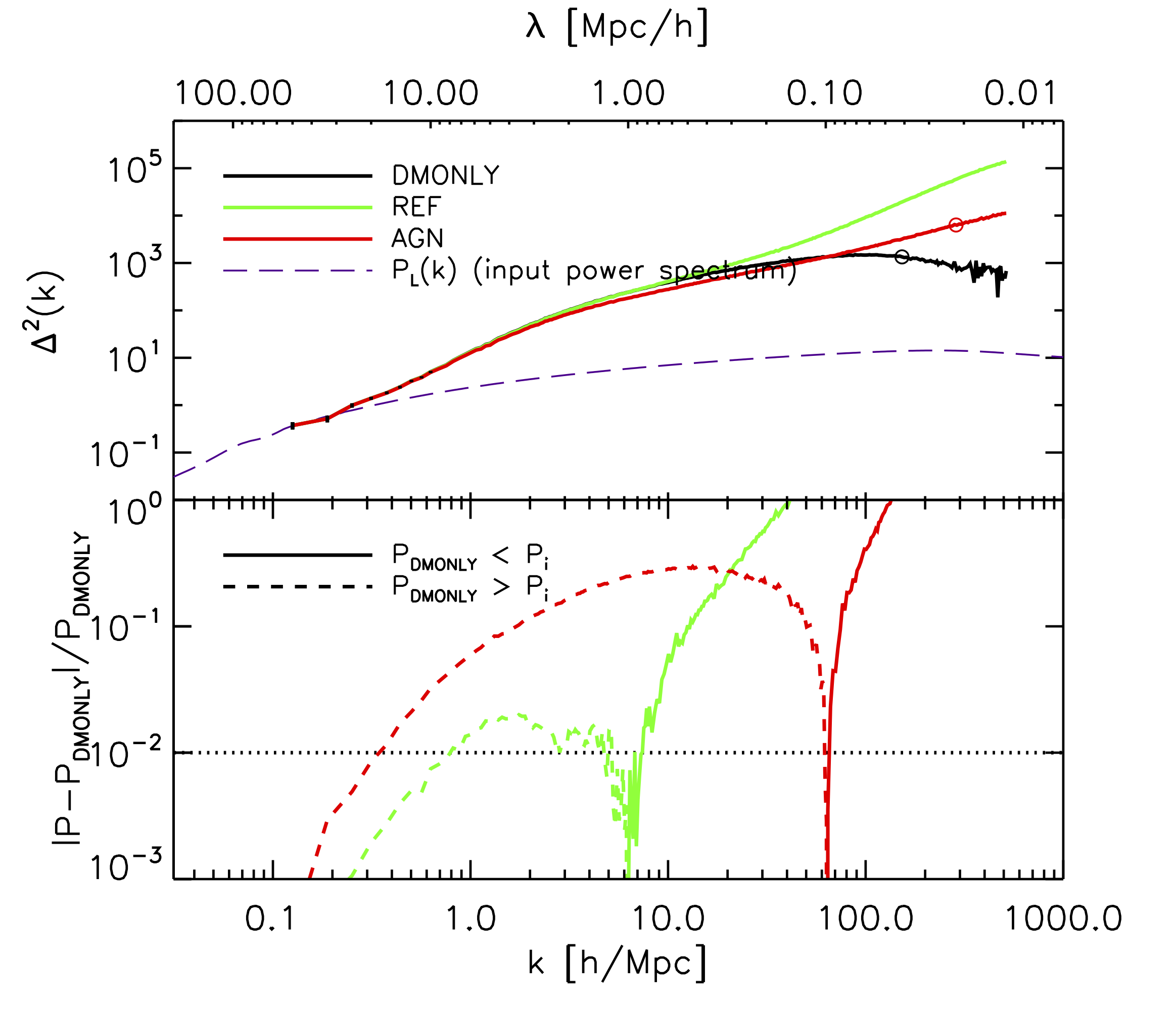}
 \caption{
 Upper panel: the total matter power spectra of REF (top solid line with highest value at $k \sim 500$), AGN (middle solid line) and DMONLY (bottom solid line), at redshift z = 0. Lower panel: the power spectrum difference between the two hydro runs and the DMONLY one; solid (dashed) curves indicate that the power is higher (lower) than for DMONLY. The dotted, horizontal line indicates the $1\%$ level. This figure is from Ref. \protect\cite{VanDaalen2011}. (note: Please refer to the online publication for a colorful figure).
}\label{fig:1}
\end{figure*}

In Figure \ref{fig:1}, they showed the dimensionless matter power spectrum $\Delta^2(k) = k^3 P(k) / 2 \phi^2$ on the upper panel and the relative difference to the DMONLY run on the lower panel. It is clear that the contribution of the baryons is significant: they decrease the power by more than 1\% for $k \sim 0.8 - 5 h Mpc^{-1}$ by comparing the DMONLY simulation with the REF simulation; the power is greatly increased at smaller scales $< 1 \Mpc (k \geq 6 h Mpc^{-1})$. The decreased power is caused by the gas pressure, which smooths the density field relative to that expected from dark matter alone. While, the increased power in the REF simulation is because radiative cooling enables gas to cluster on smaller scales than the dark matter. These results confirm the findings of previous studies, at least qualitatively \citep[e.g.][]{Rudd2008,Jing2006,Guillet2010}. However, with the AGN feedback, which is required to match observations of groups and clusters, its effect on the power spectrum is enormous: the power is reduced by $\geq 10\%$ for $k \geq 1 h Mpc^{-1}$. This could be caused by that large amounts of gas are moved to large radii due to the AGN feedback \citep[see also][]{Cui2016}. Because the AGN normally reside in massive and thus strongly clustered objects, the power is suppressed out to scales, where the removed gas can reach.

\hfill

\begin{figure*}
 \includegraphics[width=\textwidth]{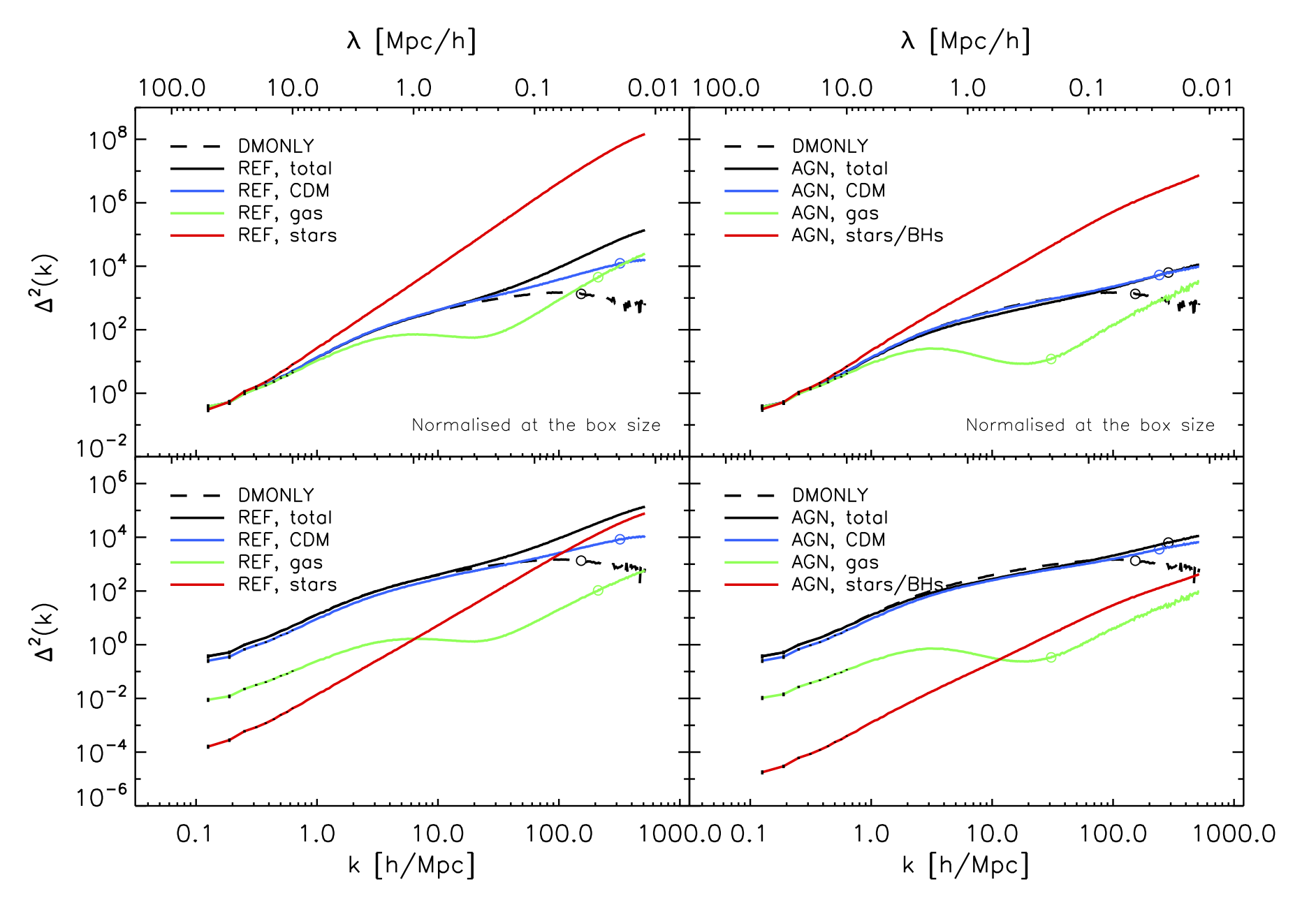}
 \caption{
Decomposing the z = 0 total power spectra into the contributions from different components. The left- and right-hand columns show results for REF and AGN. For reference, the power spectrum for DMONLY is shown with dashed black lines. This figure is from Ref. \protect\cite{VanDaalen2011}.
}\label{fig:2}
\end{figure*}

In Figure \ref{fig:2}, they showed power spectra from the REF (left panel) and AGN (right panel) simulations at z = 0. As indicated on the top left of each panel, different components are shown by different colour lines. The power spectrum for DMONLY (dashed black lines) is shown as a reference. The power spectra on top row is calculated with $\delta_i = (\rho_i - \bar{\rho_i})/\bar{\rho_i}$. This definition guarantees that all power spectra from component i converge on large scales, thus enabling a straightforward comparison of their shapes. The bottom row, on the other hand, shows the power spectra of $\delta_i = (\rho_i - \bar{\rho_{tot}})/\bar{\rho_{tot}}$, which allows one to estimate the contributions of different components to the total matter power spectrum. From the top-left panel, the baryonic components trace the dark matter well at the largest scales. However, significant differences exist for $\lambda \leq 10 \Mpc$. At scales of several hundred kpc and smaller, the difference between the CDM component (also the total component) of the reference simulation and DMONLY exceeds the change between the latter and the analytic models. This is caused by the back-reaction of the baryons on the dark matter. On the bottom left panel of Figure \ref{fig:2}, it is clear that CDM dominates the power spectrum on large scales. While the contribution of baryons is significant for $\lambda \leq 0.1 \Kpc$ and dominates below $0.06 \Mpc$. The strong small-scale baryonic clustering is a direct consequence of gas cooling and galaxy formation. For the baryonic component, the baryonic power spectrum is dominated by gas component on large scales, which has a flatter power for $\lambda \leq 1 \Mpc$ (corresponding to the virial radii of groups of galaxies) and a slightly steeper power again for $\lambda \leq 0.1 \Mpc$ (galaxy scales). While the stellar power spectrum takes control for $\lambda \leq 1 \Mpc$. The inclusion of AGN feedback greatly impacts the matter power spectrum on a wide range of scales. Comparing the top panels of Figure \ref{fig:2}, the power in both the gas and stellar components is decreased by AGN feedback for $\lambda \leq 1 \Mpc$. Through comparing the two bottom panels, the stellar power spectrum is reduced the most: about an order of magnitude on the largest scales; more than two orders of magnitude on the smallest scales. This is an expected result of the AGN feedback, which suppresses star formation, as required to solve the overcooling problem. The gas power spectrum is also dramatically dropped as a consequence of the AGN feedback. The suppression of baryonic structure by AGN feedback also makes the dominant dark matter component of the power spectrum on small scales down.

\hfill

\begin{figure*}
 \includegraphics[width=\textwidth]{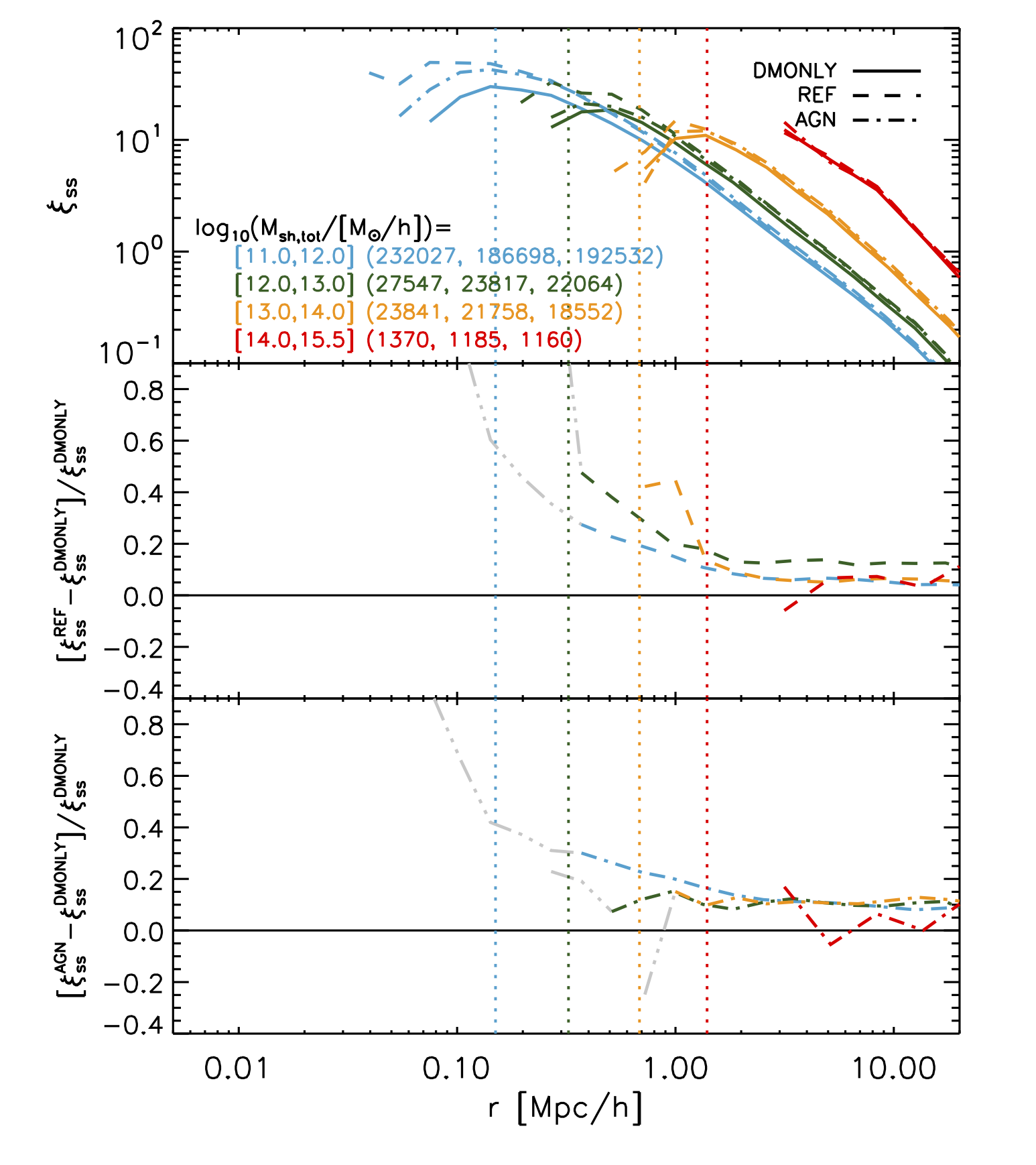}
 \caption{
Upper panel: the subhalo autocorrelation function for the three simulations: DMONLY (solid), REF (dashed) and AGN (dot-dashed lines). Different total subhalo masses results are shown with different colours, and the number of objects in each bin is indicated in the legend. The median $r_{vir}$ of the subhaloes are indicated by vertical dotted lines. Middle panel: the relative difference of subhalo clustering between REF and DMONLY. For radii, may biased due to subhalo non-detections, the curves are shown in gray. Bottom: similar to middle panel but for AGN and DMONLY. Both REF and AGN show increased clustering with a stronger effect on smaller scales. This figure is from Ref. \protect\cite{VanDaalen2014}.
}\label{fig:3}
\end{figure*}

In addition, different baryonic models investigated in \cite{VanDaalen2011} (see more details in their Figure \ref{fig:3}) showed significant changes of power spectrum at non-linear scale. It means that these baryonic models need very subtle tuning of their parameters to represent the observational results.

\hfill

\subsection{TWO-POINT CORRELATION FUNCTION}

The correlation function, $\xi(r)$, through the calculation of the excess probability to a random distribution to find the possibility of two objects at a given separation r. It is a very useful measure of the clustering of these objects as a function of scale. Comparing power spectrum, correlation can provide different views of cosmological structures. Using galaxy as a tracer, it can be used to investigate the clustering of dark matter halo \citep[for example][]{Yang2005a,Yang2005b}.

Following their work on power spectrum \cite{VanDaalen2011}, they studied the baryon effect on two-point correlation functions in \cite{VanDaalen2014} with the OWLS simulations. A parallelized brute force approach is used to calculate the correlation function. Through simple pair counts, $\xi(r)$ can be easily expressed as:
\begin{equation}
  \xi_{XY}(r)= \frac{DD_{XY}(r)}{RR_{XY}(r)} − 1.
\end{equation}
Here, X and Y denote two (not necessarily distinct) sets of objects (e.g. subhaloes and particles or haloes and haloes), $DD_{XY} (r)$ is the number of unique pairs consisting of an object from set X and an object from set Y separated by a distance r, and $RR_{XY} (r)$ is the expected number of pairs at this separation if the positions of the objects in these sets were random.

Subhalos from their simulations are identified by the SUBFIND algorithm \citep{Springel2001,Dolag2009} inside Friends-of-Friends haloes. Interested readers refer to Ref. \cite{Knebe2013} for the comparison of different subhalo finding codes, as well as the effect from the included baryonic models. Top panel of Figure \ref{fig:3} shows the subhalo autocorrelation function, $\xi_{SS}(r)$, for three different simulations: DMONLY, REF and AGN. Different colors indicate different subsamples, selected by the total mass of the subhaloes, Msh,tot. The median virial radii of subhaloes in each mass bin are indicated by vertical dotted lines. These radii are similar to the scales at which the subhalo correlation functions for DMONLY turn over. It is clear that subhalo clustering in the dark-matter-only simulation behaves quite differently from that in the baryonic models, especially on small scales ($r \leq 1 \Mpc$).

The middle and bottom panels show the relative 2-PCF difference between REF (middle)/AGN (bottom) and DMONLY simulation. All subhaloes in the baryonic simulations are typically $\sim 10\%$ more strongly clustered on large scales than their dark-matter-only counterparts. This difference is due to the reduction of subhalo mass caused by baryonic processes. For the larger subhaloes, $10^{13} < M_{sh, tot} [M\odot/h] < 10^{14}$, this offset is somewhat larger when AGN feedback is included, because supernova feedback alone cannot change the subhalo mass by as much as it can for lower halo masses \citep{Velliscig2014}. The differences between the baryonic and dark-matter-only simulations increase rapidly for $r < 2 r_{vir}$, at least for $M_{sh,tot} < 10^{14} M\odot /h$. Subhaloes from the REF simulation show significant larger clustering signal on small scales than from the AGN simulation. This seems to contradict to the results from the previous section. This is because at fixed mass range, subhaloes from the AGN simulation are less compact compared with these from the REF simulation. Due to the additional form of feedback in the AGN run, more material from the centre are pushed into outer radii, which results in a lower concentration. Similar to the subhalo 2-PCF, the galaxy 2-PCFs ($\xi_{gg}(r)$) are very similar between REF and AGN at smaller galaxy mass bins. However, it is worth to note that there is a significant difference at the largest halo mass bin, which is shown in \cite{VanDaalen2014}.

\begin{figure*}
 \includegraphics[width=\textwidth]{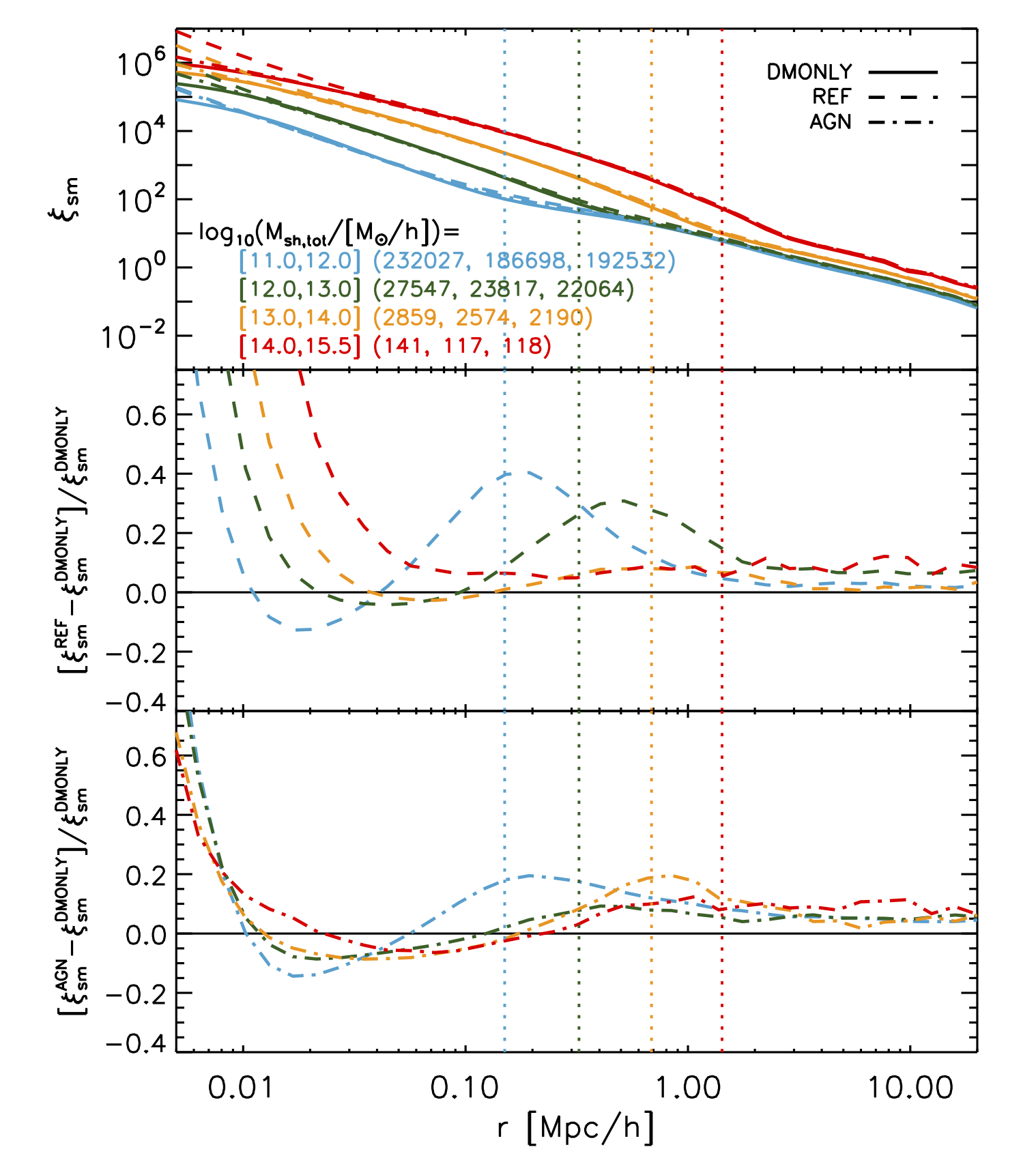}
 \caption{
As Figure \ref{fig:3}, but now for the subhalo-mass cross-correlation function, $\xi_{sm}(r)$. This figure is from Ref. \protect\cite{VanDaalen2014}.
Through these comparisons, the major reason for the increased clustering in the hydro-dynamical simulations is the lowering of the mass of objects due to galaxy formation with strong feedback. However, secondary effects, such as the resulting changes in the dynamics and density profiles of haloes, are also expected to be significant. Interestingly, \protect\cite{Despali2017} find that the presence of baryons reduces the number of subhaloes, especially at the low mass end, by different amounts depending on the model. The variations in the subhalo mass function are strongly dependent on those in the halo mass function, which is shifted by the effect of stellar and AGN feedback. We will investigate these effects on the halo mass function in Section 2.3.
}\label{fig:4}
\end{figure*}

Figure \ref{fig:4} shows the subhalo-mass 2-PCF, $\xi_{sm}(r)$ on the upper panel; the fractional difference between $\xi^{REF}_{sm}(r)$ and $\xi^{DMONLY}_{sm}(r)$ on the middle panel; the fractional difference between the $\xi^{AGN}_{sm}(r)$ and $\xi^{DMONLY}_{sm}(r)$ on the bottom panel. Again, subhaloes are generally more strongly clustered with matter in the REF and AGN than in DMONLY for scales $r > r_{vir}$. There is also a constant $\sim 5\%$ difference in favour of both REF and AGN simulations on large scales, regardless of subhalo mass. The largest differences can be up to 40 \% (20 \% for AGN) higher on intermediate scales for the lowest mass subhaloes. If sufficiently small scales are considered, this difference can be much higher for any subhalo mass. The AGN run does show a stronger decrease in clustering up to scales $r \sim  0.1 \Mpc$. While the $\xi_{sm}(r)$ at smaller mass bins from REF also show similar decrease. It is worth to note the strongly non-monotonic changes of the subhalo-mass 2-PCF between the two baryonic runs and the DMONLY one. This can be caused by the interplay between the changes in both the total subhalo mass and its mass profile. On the one hand, the lowered halo masses in the baryonic simulations tend to increase clustering at fixed mass on all scales. On the other hand, galaxy formation dissipates smoothed gas component and causes the inner halo profile to steepen (increasing clustering on small scales); the associated feedback causes the outer layers of the halo to expand (decreasing clustering on intermediate scales). These conclusions are proved in Ref. \cite{VanDaalen2014} through the 2-PCF $\xi_{sm}(r)$ that have been linked between a baryonic simulation and DMONLY which are selected based on their mass in the latter. This procedure removes the effects of changes in the subhalo masses, leaving only the effect on the mass profiles and the changes in the positions of the subhaloes.

\hfill

\subsection{HALO MASS FUNCTION}

Different to the power spectrum and 2-PCF, HMF shows another interesting statistic of the large-scale structure. Located on the central structure which connects theory with observation, HMF provides the statistical view of the halo abundance. The two most common methods used for halo identification in simulations are the FoF algorithm \citep[e.g.][]{Davis1985} and the spherical overdensity (SO) algorithm \citep[e.g.][]{Lacey1994}. Interested readers refer to Ref. \cite{Knebe2011} for the comparison of different halo finding codes.

A series of three versions of cosmological simulations are used in Ref. \cite{Cui2012a,Cui2014a} for their study. Starting from the same initial condition, these simulations share the same number of dark matter particles ($1024^3$) and gas particles ($1024^3$) within a simulation box size of 410 $ \Mpc$. A first hydro-dynamical simulation includes radiative cooling, star formation and kinetic SN feedback (CSF hereafter), while the second one also includes the effect of AGN feedback (AGN1 hereafter). As for the DM simulation, it simply replaces the gas particle by collisionless particles, so as to have the same description of the initial density and velocity fields as in the hydro-dynamical simulations.

\begin{figure*}
 \includegraphics[width=0.5\textwidth]{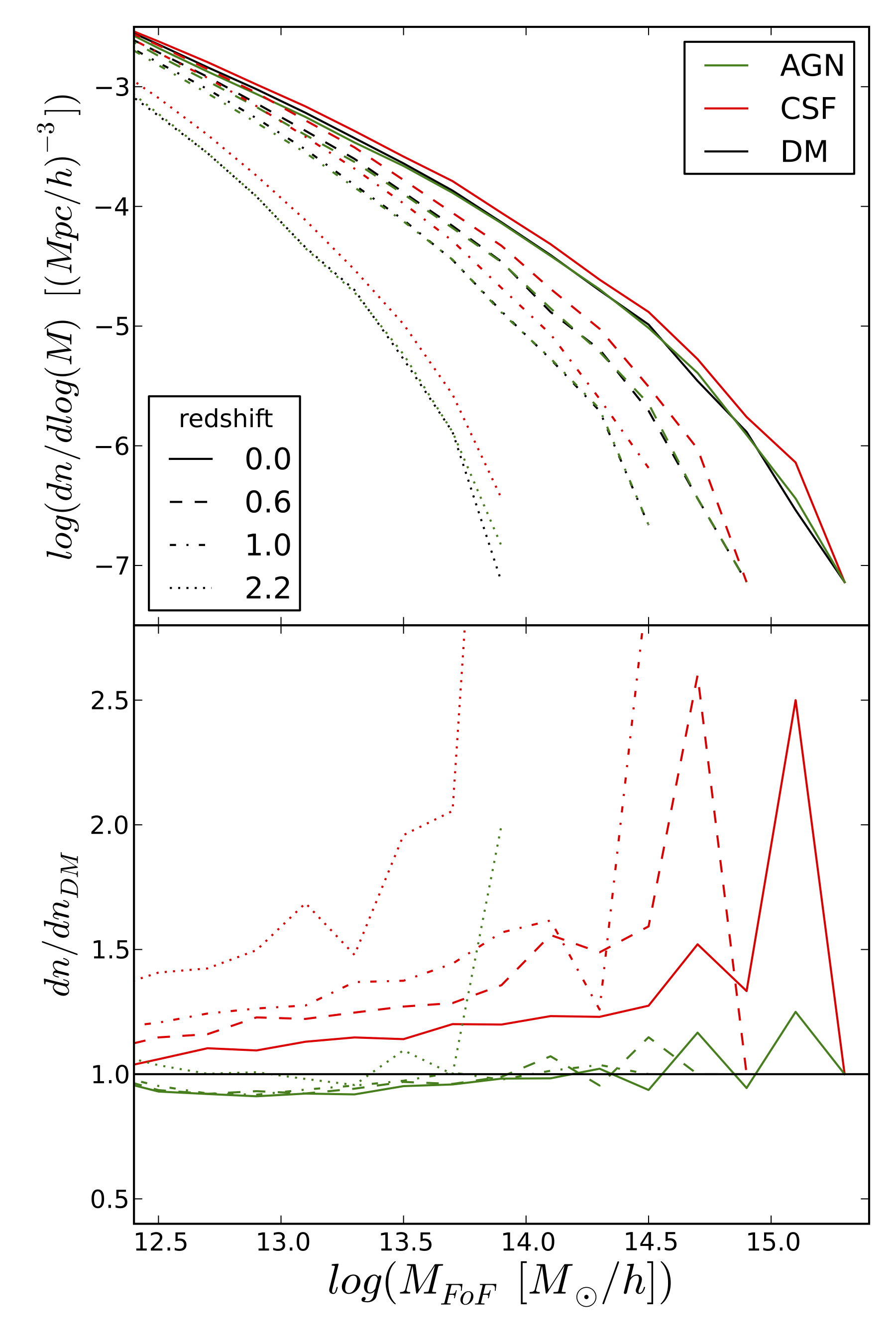}
 \caption{
Upper panel: FoF HMFs from DM, CSF and AGN. The HMFs from CSF are always higher than the results from AGN and DM. Different redshifts are shown with different line styles (see the lower left legend for details). Lower panel: relative difference between the HMFs from the hydro-dynamic simulations and from the DM simulation from all four redshifts. This figure is from Ref. \protect\cite{Cui2014a}. (Please refer to the online publication for a colorful figure).
}\label{fig:5}
\end{figure*}

FoF HMFs are compared on the top panel of Figure \ref{fig:5} between the three different versions of simulations. While the bottom panel shows the halo number ratio in a mass bin respected to the DM simulation. The baryonic effect from the CSF with respect to the DM case has clear redshift evolution as well as halo mass dependence. From higher redshift to lower redshift, the HMF ratios between the CSF and DM runs decrease from $\sim $1.6 to $\sim $1.1, with a weak increasing trend along halo mass changes. Quite remarkably, including AGN feedback in the baryonic model reduces the difference with respect to the DM-only case: the HMF ratio drops to about unity for massive haloes with $M_{FoF} \approx 10^{14} \hMsun$, while at smaller halo mass it decreases to $\sim $0.9 for $M_{FoF} \approx 10^{13} \hMsun$. Different to the CSF case, there is no clear redshift evolution in these ratios from z = 1 to 0. At the highest redshift, z = 2.2, this HMF ratio keeps fluctuating around 1. This could be a consequence of the limited statistics of haloes due to the finite box size.

\hfill

Using the PIAO\footnote{https://github.com/ilaudy/PIAO} code \citep{Cui2014}, the SO haloes are identified with three overdensities $\Delta_c = 2500, 500$ and 200. These HMFs are shown in Figure \ref{fig:6} from left to right top panels, respectively. While the HMF ratios from the CSF and AGN simulations with respect to the one from DM run are shown in lower panels. Baryons show a larger impact on the HMF at the higher overdensity. With $\Delta_c$ = 2500, the ratio between the CSF and DM HMFs shows a redshift evolution ranging from $\sim $1.4 at z = 0 to $\sim $2.5 at z = 2.2, but with no significant dependence on the halo mass. At lower overdensities, the redshift evolution becomes weaker and the differences with respect to the DM case are also reduced. When AGN feedback is included in the hydro-dynamical simulation, the corresponding HMF drops below the HMF from the DM simulation, by an amount that decreases for lower $\Delta_c$ values, with no evidence for redshift dependence on the HMF difference. Generally speaking, the baryonic effect on the HMF goes in the same direction, qualitatively independent of whether FoF or SO halo finders are used. However, as expected, quantitative differences between FoF and SO results are found, especially for the AGN case. This is rooted in intrinsic algorithm difference of these halo finder methods (we refer interested readers to Ref. \cite{Knebe2011} for details).

\begin{center}
\begin{figure*}
 \includegraphics[width=\textwidth]{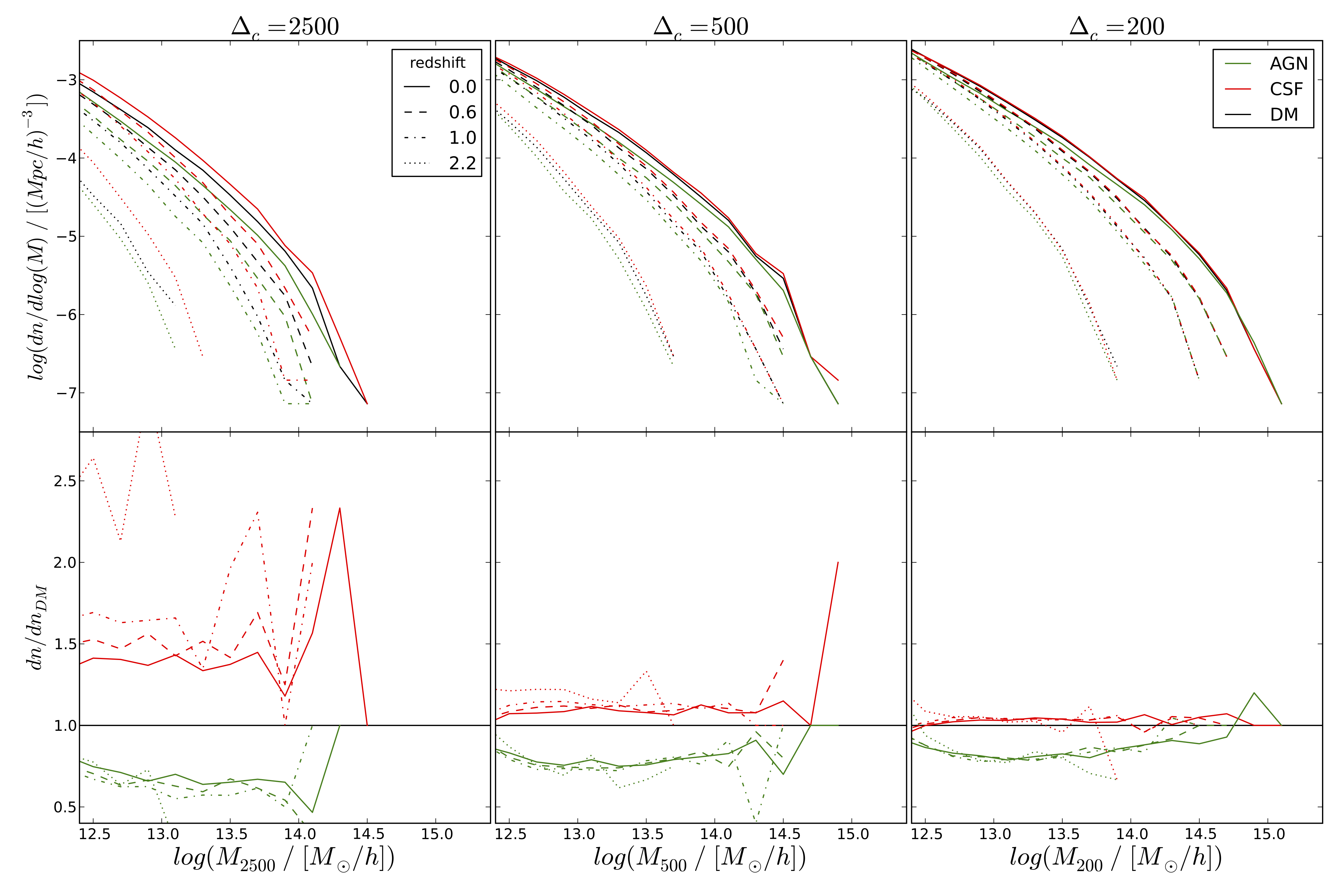}
 \caption{
Similar to Figure \ref{fig:5} but for the HMFs of SO haloes. Three different overdensities $\Delta_c$ = 2500 (left-hand panel), 500 (middle panel) and 200 (right-hand panel) are used to identify SO haloes. Again, top panels show the SO HMFs, while bottom panels show the HMF difference. This figure is from Ref. \protect\cite{Cui2014a}.
The three simulations share the same dark matter particles, which have the same progressive identification number (ID). Therefore, we can use the halo from the DM simulation as the reference. The halo in the CSF or AGN simulation is defined as the counterpart of the DM halo, if it includes the largest number of DM particles belonging to the latter. In their paper, a matching rate is defined as the ratio of matched to total number of dark matter particles in the DM halo. To avoid multiple-to-1 matching from CSF/AGN simulation to the DM one, only haloes with matching rate larger than 0.5 are selected.
}\label{fig:6}
\end{figure*}
\hfill
\end{center}

Figure \ref{fig:7} shows the halo mass ratios between these matched haloes. Red points indicate the halo pairs, which are coming from CSF and DM simulations, while green points are for the pairs from AGN and DM simulations. The thick lines show the mean value of these data points computed within each mass bin (magenta for CSF and blue for AGN, respectively). For the CSF-DM halo pairs, the increased halo mass is almost independent of redshift. At each redshift, the ratio shows a weak decrease with halo mass, from $\sim$ 1.1 at $M_{500} = 10^{12.5} \hMsun$ to $\sim$ 1.05 at $M_{500} > 10^{13.5} \hMsun$, then becoming constant. However, for the AGN-DM pairs, the strong AGN feedback makes the ratio go in the opposite direction (decreased halo masses). Thereby, this will result in a decreased HMF, which has been shown in Figure \ref{fig:6}. There also shows no evidence of redshift evolution for the halo mass ratio, at least below z = 1.0. However, this ratio shows a strong dependence on the halo mass, which increases from $\sim$ 0.8 at $M_{500} = 10^{12.5} \hMsun$ to $\sim$ 1 for the most massive haloes found in their simulation box.

\begin{center}
\begin{figure*}
 \includegraphics[width=\textwidth]{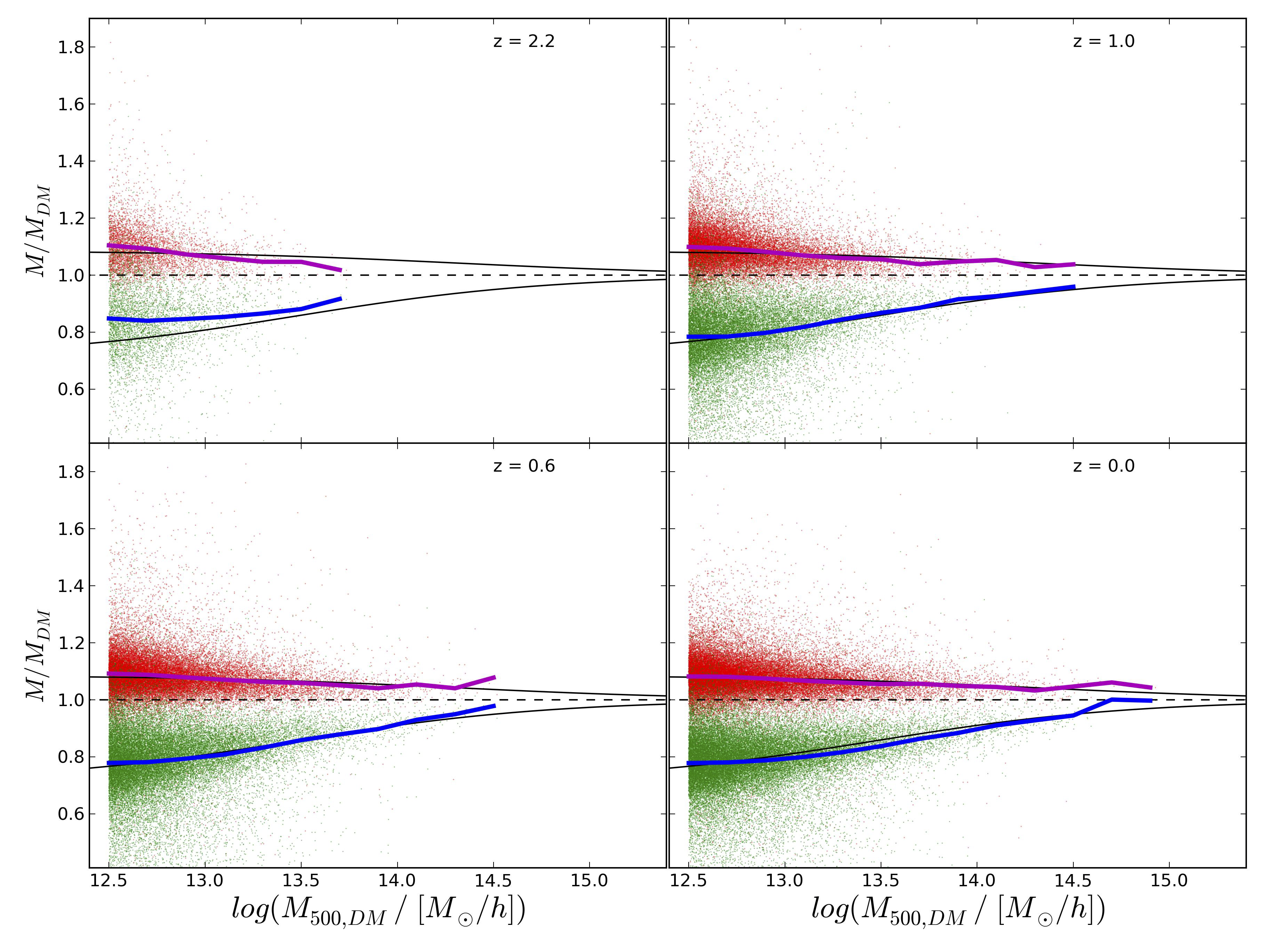}
 \caption{
The ratio of masses of matched SO haloes as function of MDM computed for $\Delta_c$ = 500 at four different redshifts. Each data point indicates a halo mass ratio between the matched CSF or AGN halo to its corresponding DM one. These misty data points above the horizon dashed lines are normally from CSF run, while lower ones are coming from AGN run. The mean values of these ratios within each mass bin are shown by thick magenta (CSF) and blue lines (AGN), respectively. The solid black lines are the best-fitting results for the mass correction, which are used for the HMF correction in Ref. \protect\cite{Cui2014a}. This figure is from Ref. \protect\cite{Cui2014a}. (Please refer to the online publication for a colorful figure).
}\label{fig:7}
\end{figure*}
\hfill
\end{center}

\subsection{Large-scale Environments}

On even larger scales, matter in the Universe can be roughly distributed into knots, filaments, sheets and voids, which form the rather prominent cosmic web seen both in numerical simulations of cosmic structure formation and observations of the distribution of galaxies. These four different cosmological structures are a natural outcome of gravitational collapse. A detailed understanding of the large-scale environment helps us to model both how the dark matter or galaxies are distributed and evolve from early times to the present day. However, dark-matter-only simulations are normally used with aforementioned methods to study the properties of the large-scale structure \citep[e.g.][]{Hahn2007, Lee2008, Zhang2009, Metuki2015,Yang2017} -- which is dominated by the effects of gravity and hence dark matter. Although, it is well known that baryons mainly impact upon structure formation on small, non-linear scales \citep[see][and the references therein]{Cui2016}. It therefore remains unclear whether these cosmological structures at large scale are affected by the baryons or not.

Using the same cosmological simulations in \cite{Cui2012a, Cui2014a}, they applied the Hessian matrix methods -- both on velocity shear tensor \citep[{\sc Vweb}][]{Hoffman2012} and on the tidal field of the gravitational potential \citep[{\sc Pweb}][]{Hahn2007} to classify these cosmological structures. These three sets of simulations allow them to make a solid statement of the influences of baryons on these cosmological structures. We refer interesting readers to \cite{Cui2017b} for the detailed analysis.

Fig.~\ref{fig:be} shows both the volume fraction (indicated by subindex V in the symbol names) and the mass fraction (indicated by subindex M) for cells of a given cosmological structure (which is listed on the x-axis). The cosmological structures identified with {\sc Vweb} are shown in the left column, while right column is for the cosmological structures identified with the {\sc Pweb} code. As shown in the legend on the top left panel, different colour symbols represent different runs. For both methods, we see excellent agreement between these three simulation runs for both volume and mass fractions. The mean matter density decreases from knot to void regions. Therefore, it is not surprising to see that their volume fractions are different to their mass fractions.

The lower panel of Fig.~\ref{fig:be} shows the quantitative difference between the respective fractions runs with respect to the DM run. On the lower left panel, the AGN run tends to have a slightly larger ($\sim 2$ per cent) volume fraction in the knot region, while the CSF run gives $\sim$ 1 per cent lower volume fraction than the DM run. However, there is almost no change of the mass fractions. Without AGN feedback, the CSF run tends to have more concentrated knots with a relatively weaker velocity field, which tends to occupy less spatial volume; while the strong AGN feedback not only stops star formation but also pushes matter into outer regions, which results in a large volume with a higher velocity field \citep[see discussion in][]{Ragone-Figueroa2013, Cui2014a, Cui2016a}. In the void regions, the AGN run tends to have a larger mass fraction ($\sim 2$ per cent), while the CSF run tends to have a lower mass fraction ($\sim 1$ per cent) compared to the DM run. The mass change may be caused by matter distribution in this region --- more matter is expelled due to AGN feedback. In filament and sheet regions, both mass and volume fractions show almost no change between these three runs. As the {\sc Pweb} code directly uses the second derivatives of the potential, which are directly connected to the density via Poisson's equation, to classify these structures, there is even less difference ($\lesssim 1$ per cent) between the two hydrodynamical runs and the DM simulation for all cosmological structures.

\begin{center}
  \begin{figure*}
   \includegraphics[width=\textwidth]{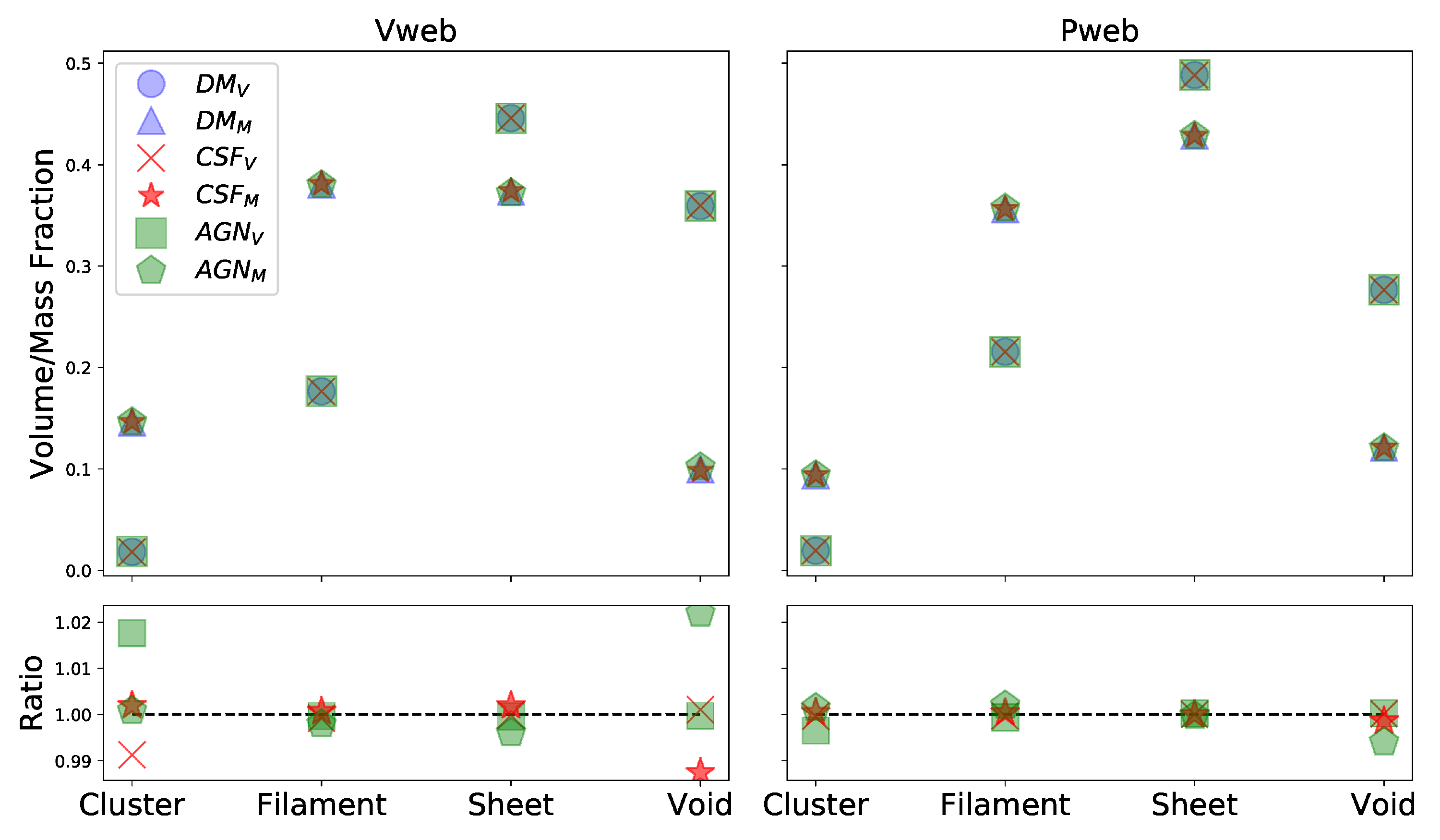}
   \caption{
    Upper panel: the volume and mass fractions of these cosmological structures. Lower panel: respective ratios between the CSF/AGN and DM runs. Left column shows the results from the {\sc Vweb} code, while the right column is for the {\sc Pweb} code. The meaning of the symbols is shown in the legend on the top left panel where sub-indexes V and M indicate volume and mass fractions, respectively. This figure is from Ref. \protect\cite{Cui2017b}.}\label{fig:be}
  \end{figure*}
\hfill
\end{center}

Besides the finding that these cosmological structures are insensitive to the uncertain knowledge of baryonic processes, \cite{Cui2017b} also investigated the cosmological structures identified with only baryonic matter -- gas component. They found that the gaseous structures, especially filaments and sheets, are almost identical for both methods to the structures identified with only dark matter component. This potentially allow us to use the cosmic web from up coming SKA observations to constrain cosmology models bias-free.

\subsection{SUMMARY}

Stepping from dark-matter-only to hydro-dynamical simulations allows us to view the galaxy formation and evolution in the Universe in a self-consistent and realistic way. Hydro-dynamical simulations estimate tight connections between theoretical and observational researches, therefore providing a perfect test lab for examining theories. Through these comparisons between state-of-the-art hydro-dynamical simulations and dark-matter-only ones, we summarized the recent findings of baryonic effects on the large-scale structure of the Universe by showing the changes on power spectrum, two-point correlation function and halo mass functions:

\begin{itemize}
  \item {\em Power spectrum.} There is a decreased power (1\%) at $k \sim 0.8 - 5~ h Mpc^{-1}$ ($\sim 8 - 1 ~\Mpc$). At smaller scales (< 1 $\Mpc$ or $k > 6 ~h Mpc^{-1}$), the power rises quickly far above the dark-matter-only simulations because of the baryon processes. However, this increase is reduced by >10\% when the AGN feedback is switched on. Power spectra for individual component reveal at which scales they are responsible for these changes: cold dark matter dominates the power spectrum on large scales; gas component contributes mildly over all scales; stellar component is the reason for the high power at small scales.
  \item {\em Two-point correlation function.} The correlation functions for subhalo are typically $\sim$ 10\% higher in hydro-dynamical simulations than in dark-matter-only ones. While this change is significantly larger at smaller scale. With AGN feedback on, the differences are slightly higher at large scale and lower at small scale compared to the reference one without AGN feedback. Subhaloes are also strongly clustered with matter in the baryonic simulations than in the dark-matter-only ones.
  \item {\em Halo mass function.} The halo mass functions are also higher from hydro-dynamical simulations than from dark-matter-only simulations. These differences depend on redshifts, halo mass ranges and halo finding methods. With AGN feedback, the halo mass functions are normally lower than their counterparts from the dark-matter-only simulations. These changes are vividly indicated by the variances of the halo masses, which are matched one to one between these simulations.
  \item {\em Cosmological structures.} On larger scale, in agreement with the other findings, the cosmological structures, i.e. knots, filaments, sheets and voids, are almost unaffected by the baryonic models.
\end{itemize}

Besides these statistics methods investigated in upper paragraphs, baryonic processes can also leave an impact on cosmological structures, such as cosmic webs, sheets and voids. Using the EAGLE simulation, \cite{Paillas2017} studied the effect of baryons on void statistics. They found that the dark-matter-only simulation produces 24\% more voids than the hydro-dynamical one, but this difference comes mainly from voids with radii smaller than 5 Mpc. This contradicts to the finding in \cite{Cui2017b}. These difference can be caused by 1.) the identification methods; 2) the simulation volume; 3) the detailed baryonic models.

However, at non-linear scale, all these results strongly depend on the included baryonic models and simulation codes. There is no guarantee of a perfect model yet, especially that most of the implanted baryonic models are based on observational relations. Starting from the same initial condition of a galaxy cluster, the recent nIFTy project \citep{Sembolini2016,Sembolini2016a,Elahi2016,Cui2016,Arthur2017} has made vast comparisons between different simulation codes as well as baryonic models included in them. There is a good agreement between these simulation codes for the dark-matter-only runs. A larger disagreement is shown between the classic SPH codes and mesh/modern SPH/moving mesh codes for the non-radiative hydro-dynamical runs. In the full physics runs, the largest difference is lying between the runs with AGN feedback and the ones without AGN feedback. However, even inside both families, there are a lot of variances between different simulation codes.

To simulate the observed Universe, more efforts are needed to understand the sub-grid baryonic models, such as their parameter choice, resolution and method dependence. To understand and pin down these sub-grid models, we need direct and detailed comparisons between the simulation results and observational ones. Thus, a one-to-one comparison is much helpful than the statistical relations. These constrained simulation projects aiming to represent the observed Universe, the ELUCID project \citep{Wang2013,Wang2014,Tweed2017,Wang2016a,Wang2017}, the CLUES project \citep[][etc.]{Gottloeber2010,Sorce2014,Carlesi2016} and the APOSTLE project \citep{Sawala2016}, point to the direction of future simulation studies.

\section{Acknowledgements}

The authors are particularly grateful to Dr. Marcel van Daalen for freely using the figures from his papers \cite{VanDaalen2011} and \cite{VanDaalen2014}. WC would like to thank his wife Ms Yufang Liu for her kind help. He also acknowledges the supports from the {\it Ministerio de Econom\'ia y Competitividad} and the {\it Fondo Europeo de Desarrollo Regional} (MINECO/FEDER, UE) in Spain through grant AYA2015-63810-P as well as the Consolider-Ingenio 2010 Programme of the {\it Spanish Ministerio de Ciencia e Innovaci\'on} (MICINN) under grant MultiDark CSD2009-00064.
Y.Z. acknowledges the support from the 973 Program (No. 2015CB857002).

\bibliographystyle{mnras}
\bibliography{baryon-effect}

\bsp
\label{lastpage}
\end{document}